\documentclass[final,nofootinbib,pre,twocolumn,showpacs,showkeys,groupaddress,preprintnumbers,floatfix]{revtex4-1}

\usepackage{dynlearn}

\renewcommand{\H}{\operatorname{H}}

\newcommand{\cs}{s}                
\newcommand{\CQST}{\mathbb{T}}       

\newcommand{\MSym}{\MeasSymbol}
\newcommand{\msym}{\meassymbol}

\newcommand{\Abet}{\ProcessAlphabet}
\newcommand{\SSet}{\CausalStateSet}

\newcommand{\MxSSet}{\AlternateStateSet}
\newcommand{\MxSMeasure}{\mu}
\newcommand{\MxSDyn}{\mathcal{W}}
\newcommand{\mxst}{\eta}

\newcommand{\One}{ {\mathbf{1} } }

\newcommand{\CmuDim}{d_{\mu}}
\newcommand{\LCEDim}{d_\text{LCE}}
\newcommand{\BoxDim}{d_\text{BC}}

\begin{document}

\def\ourTitle{
Measurement-Induced Randomness and Structure\\
in Controlled Qubit Processes
}

\def\ourAbstract{
When an experimentalist measures a time series of qubits, the
outcomes generate a classical stochastic process. We show that measurement
induces high complexity in these processes in two specific senses: they are
inherently unpredictable (positive Shannon entropy rate) and they require an
infinite number of features for optimal prediction (divergent statistical
complexity). We identify nonunifilarity as the mechanism underlying the
resulting complexities and examine the influence that measurement choice has on
the randomness and structure of measured qubit processes. We introduce new
quantitative measures of this complexity and provide efficient algorithms for
their estimation.
}

\def\ourKeywords{
  stochastic process, hidden Markov model, \texorpdfstring{\eM}{epsilon-machine}, causal states, mutual information
}

\hypersetup{
  pdfauthor={James P. Crutchfield},
  pdftitle={\ourTitle},
  pdfsubject={\ourAbstract},
  pdfkeywords={\ourKeywords},
  pdfproducer={},
  pdfcreator={}
}

\title{\ourTitle}

\author{Ariadna E. Venegas-Li}
\email{avenegasli@ucdavis.edu}

\author{Alexandra M. Jurgens}
\email{amjurgens@ucdavis.edu}

\author{James P. Crutchfield}
\email{chaos@ucdavis.edu}

\affiliation{Complexity Sciences Center and Physics Department,
University of California at Davis, One Shields Avenue, Davis, CA 95616}

\date{\today}
\bibliographystyle{unsrt}

\begin{abstract}
\ourAbstract
\end{abstract}



\preprint{\arxiv{1908.XXXX}}

\date{\today}
\maketitle

\setstretch{1.1}

\paragraph*{Introduction}
\label{sec:intro} 
Temporal sequences of controlled quantum states are key to fundamental physics
and its engineering deployment. Quantum entanglement \cite{Schr35b} between
emitted qubits, such as photons \cite{Wu50a}, is central to Bell probes
\cite{Bell64a} of inconsistencies between quantum mechanics and local hidden
variable theories \cite{Eins35a}. Complementing their scientific role,
entangled qubits are now recognized as basic resources for quantum
technologies---quantum key distribution \cite{Eker91a}, teleportation
\cite{Benn93a}, metrology \cite{Giov06a}, and computing \cite{Raus01a}. The
quest there is for qubit sources that allow \emph{on-demand} generation: at a
certain time a source should emit one and only one pair of entangled photons.
Qubit sources should also be \emph{efficient}: qubits emitted and collected
with a high success rate. And, individual qubits should have \emph{specified
properties}. In EPR experiments photons in emitted pairs should be identical
from trial to trial. And, in communication systems polarization states should
manipulable at the highest possible rates \cite{Land19a}.

Much experimental effort has been invested to develop qubit sources that, for
example, extract entangled photons from trapped atoms \cite{Free72a,Mill05a},
spontaneous parametric down-conversion \cite{Kwia95a}, quantum dots
\cite{Bens00a}, and related CQED systems \cite{Walt06a}. To date, though, there
is still no single qubit source that exactly meets the performance desiderata.
The on-demand criterion has been particularly vexing \cite{Eisa2011a}.
Addressing these challenges leads rather directly to a common question, one that
touches on both fundamental physics and quantum engineering, How to
characterize the statistical and structural properties of qubit time series?
The underlying challenge is that a systematic description of quantum processes
with memory in terms of experimental measurements has yet to be given
\cite{Poll18a}.

To address these challenges we first introduce qubit information sources
observed through quantum measurement. Noting that their outputs---what an
experimentalist sees---are classical stochastic processes, we review measures
of complexity for the latter. This leads directly to complexity measures
appropriate to qubit processes and to our identifying the mechanism in quantum
measurement that generates the observed complexity.

We concern ourselves with qubit sources that generate single qubits at a time,
putting aside entangled pairs for now, and we ask the source to determine which
qubit property, out of a finite set, is generated at each time. We imagine that
the on-demand source is used repeatedly, producing an arbitrarily-long time
series, which we call a \emph{qubit process}. A simple example arises when
monitoring sequential emissions from a blinking quantum dot
\cite{Gall11a,Efro16a}. Qubit processes are our main object of study; we refer
to their generators as \emph{controlled qubit sources} (CQSs). We ask how
random and structured they appear to an experimentalist. (SM
\ref{qpp:QuantumFormalism} highlights the features of the quantum formalism we
use.)

\begin{figure}[htbp]
\centering
\includegraphics[width=\columnwidth]{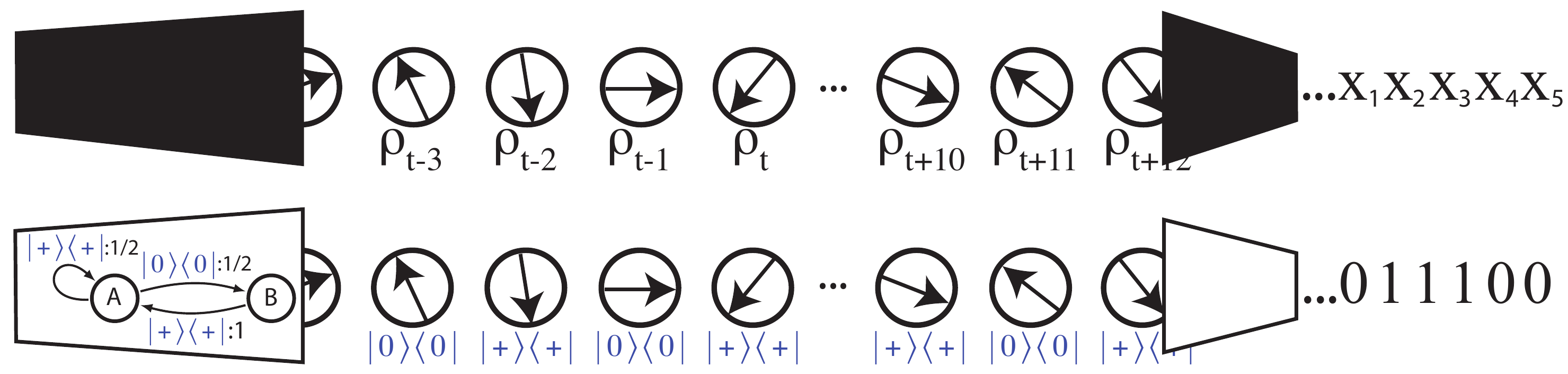}
\caption{Two kinds of qubit source: (Top) A general controlled qubit source
	(CQS) as a discrete-time quantum dynamical system, represented by a black
	box, generates a time series of qubits $\rho_{t-2} \rho_{t-1} \rho_t
	\ldots$.  (Bottom) A classically-controlled qubit source (cCQS) generates a
	qubit process $\ket{0}\bra{0} \ket{+}\bra{+} \ldots$. Measuring each qubit,
	an observer sees a classical stochastic process: (Top) $\ldots x_1 x_2 x_3
	x_4 x_5$, (Bottom) $\ldots 011100$. What can we learn from the classical
	process about the underlying cCQS?
	}
\label{fig:CQSsetup}
\end{figure}

To make headway analyzing qubit process complexity we introduce a qubit source
that is classically controlled and classically measured---for short,
\emph{classically-controlled qubit source} (cCQS). That is, the qubit generator
proper is sandwiched between controller and measurement apparatus, both built
out of classical observables. The controller determines the kind of qubit
generated at each time. Figure \ref{fig:CQSsetup} (top) illustrates the setup:
a black box, representing a quantum system, emits a qubit in quantum state
$\rho_t$ at each time $t$. More concretely, the lower panel shows an example,
revealing that the controller inside the black box is a finite-state hidden
Markov model (HMM) that emits qubits in various quantum states.

We restrict the qubit states emitted by the cCQS to be pure-state density matrices; that is,
$\rho^2 = \rho$. This limits the type of correlations that can be present
across the qubit time series to classical correlations and leaves time series
with temporal entanglement for future exploration. Since each qubit $\rho_t$
is in a pure state, the quantum state of the random variable chain that forms
the time series can be regarded as the tensor product of the individual qubits:
$\ldots \rho_{t-2} \otimes \rho_{t-1} \otimes \rho_{t} \ldots$.

This simple setup raises several natural questions about characterizing qubit processes generated by CQSs. How random is the qubit process? How much memory does the source use to generate the qubit series? Can we identify the internal control mechanism from the qubit time series alone?

\begin{figure*}
\centering
\includegraphics[width=\textwidth]{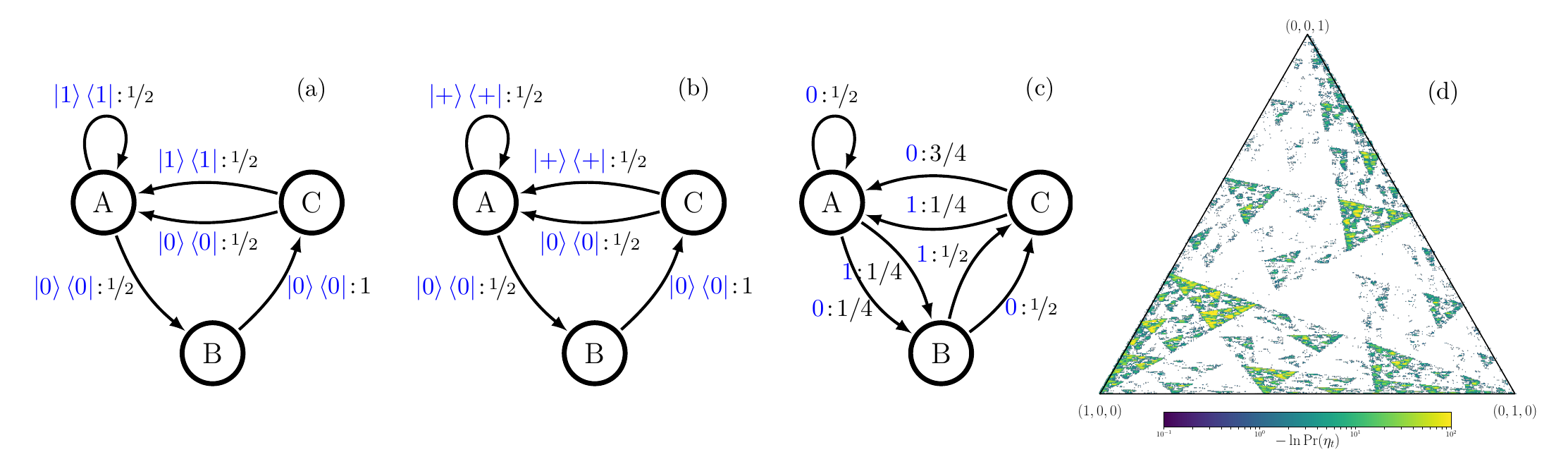}
\caption{
 	(a) Three-state classically-controlled qubit source (cCQS) that
	generates a process consisting of orthogonal qubits: 
 	(i) controller states $\CausalStateSet = \{A, B, C\}$;
 	(ii) orthogonal qubit alphabet
		$\mathcal{A}_Q = \{\rho_0 = \ket{0}\bra{0}, \rho_1 = \ket{1}\bra{1}\}$;
 	(iii) labeled transition matrices $\CQST^{\rho_0}$ and $\CQST^{\rho_1}$,
		whose state-transition probability components $(\CQST^{\rho_k})_{ij}$
		can be read from the diagram; and
 	(iv) stationary state distribution
 		$\pi = \protect\begin{pmatrix} 1/2 & 1/4 & 1/4 \protect\end{pmatrix}$.
 	If the controller is in state $A$ it has equal probabilities of staying
		in that state or transitioning to state $B$. 
 	If it transitions to state $B$, then the system emits a qubit in pure
 		state $\ket{0}\bra{0}$. 
 	In the next time step the machine must transition to state $C$ and
 		emits another $\ket{0}\bra{0}$ qubit.
		Then, in state $C$ it transitions back to state $A$, emitting either
		pure state $\ket{0}\bra{0}$ or $\ket{1}\bra{1}$ with equal probability.
 	As the cCQS runs, a time series of orthogonal qubits is generated. 
 	(b) Nonorthogonal-qubit cCQS: Three-state HMM that outputs nonorthogonal
		qubits in pure states $\ket{0}\bra{0}$ and $\ket{+}\bra{+}$, where
 		$\ket{+} = (\ket{0}+\ket{1})/\sqrt{2}$.
 	(c) HMM presentation for the classical stochastic process resulting from
 		measurement ($\theta = \pi/2$) of the quantum process generated by (b).
 	(d) Mixed states for the stochastic process generated by (c) in that HMM's
		state distribution simplex. Each mixed state is a point of the form
		$(p_A, p_B, p_C)$ with probabilities of being in state $A$,
		$B$, or $C$ of (c).
  	}
\label{fig:cCQSs}
\end{figure*}

\renewcommand{\past}{\ms{-\infty}{0}}
\newcommand{\realize}{\ms{-\infty}{\infty}}
\renewcommand{\Future}{\MS{0}{\infty}}
\renewcommand{\Past}{\MS{-\infty}{0}}

Now, if one replaces the CQS with a classical system that emits symbols $X$
taking values in a discrete alphabet ($\msym \in \Abet$), the output is a
\emph{stochastic process} $\Pr(\Past,\Future)$ over pasts $\Past$ and futures
$\Future$. Here, $X_t$ denotes the random variable at time $t$ and a block is
denoted $X_{t:t+l} = X_t, X_{t+1}, \ldots, X_{t+l-1}$. In this classical
setting, all of the questions posed above can be answered constructively. For
example, the process' \emph{Shannon entropy rate} $\hmu$:\\[-15pt]
\begin{align*}
\hmu = \lim_{\ell \to \infty} \frac{\H[\Prob(\MS{0}{\ell})]}{\ell}
\end{align*}\\[-15pt]
measures a process' randomness as the rate of increase of information in
length-$\ell$ sequences---in the \emph{Shannon block entropy}
$\H[\Prob(\MS{0}{\ell})]$.

Determining a process' memory requires analyzing a particular kind of
edge-labeled HMM generator. First, this HMM is \emph{unifilar}: for each state
$\cs_k \in \CausalStateSet$ and each symbol $\msym$ there is at most one
transition from $\cs_k$ that emits $\msym$. Second, its states are
\emph{probabilistically distinct}: For every pair of states $\cs_k, \cs_j \in
\CausalStateSet$ there exists some finite word $w = \msym_0 \msym_1 \ldots
\msym_{\ell-1}$ such that $\Pr(w|\cs_k) \neq \Pr(w|\cs_j)$. These properties
define a process' \emph{causal states}, they capture the minimal amount of
information from the past to optimally predict the process' future. Together
with their transition dynamic $\{T^{(\msym)}: \msym \in \Abet\}$ the causal
states form the process' minimal optimally-predictive model---its \emph{\eM}
\cite{Crut12a}. (Figures \ref{fig:cCQSs} (a), (b), and (c) give examples of
unifilar and nonunifilar HMMs.) Though a seemingly-innocent structural
property, we show that unifilarity plays a decisive role in quantum
measurement.

Most immediately, in the classical setting unifilarity allows one to calculate
the entropy rate directly from the \eM\ as the state-averaged symbol
uncertainty:\\[-15pt]
\begin{align}
\hmu & = - \sum\limits_{\cs \in \CausalStateSet}
  \Prob(\cs) \sum\limits_{\msym \in \Abet}
  \sum\limits_{\cs^\prime \in \CausalStateSet}
  T^{(\msym)}_{\cs \cs^\prime} \log T^{(\msym)}_{\cs \cs^\prime}
  ~.
\label{eq:uni_hmu}
\end{align}\\[-15pt]
Given that a process generates information at rate $\hmu$ bits per measurement,
one is next interested in the resources required to predict measurements. This
is given by the \emph{statistical complexity} $\Cmu$---the Shannon information
or average memory in the causal states:\\[-15pt]
\begin{align}
\Cmu & = -\sum\limits_{\cs \in \CausalStateSet}
  \Prob(\cs) \log_2 \Prob(\cs)
  \label{eq:Cmu}
  ~.
\end{align}\\[-15pt]
(See SM \ref{app:processes} and \ref{app:moc}.)

In light of these complexity metrics, we can now state our main result: Even
with a finite-state control, \emph{generically a CQS produces a measured qubit
process whose minimal predictor requires an infinite number of causal states.}
Prediction resources ($\Cmu$) diverge; though at a quantifiable rate. We
establish the result constructively, by determining $\hmu$ and exploring
$\Cmu$'s divergence for qubit processes and by identifying the driving
mechanism as measurement-induced nonunifilarity. These steps require
introducing novel concepts from ergodic theory and dynamical systems and new
efficient algorithms, whose development appears in SM \ref{app:MixedStates} and
\ref{app:CmuDim}.

\paragraph*{Qubit Processes}
\label{sec:QProcess} 

Generating the qubits in a time series is governed by a cCQS that, without loss
of generality, we take to be an \eM\ for which the symbols emitted during
state-to-state transitions consist of qubit-states. This choice ensures that the
source's internal complexity used in generating the qubit process can be
quantified. Both the entropy rate $\hmu^g$ and the statistical complexity
$\Cmu^g$ of the generating, internal-state process can be exactly computed.

The states of the qubits output by a cCQS form a stochastic process;
two examples of the latter
are shown in Figs. \ref{fig:cCQSs} (a) and (b). The caption explains how
these qubit-generating state machines operate. The cCQS in (a) generates a
qubit time series of orthogonal pure states $\ket{0}\bra{0}$ and
$\ket{1}\bra{1}$. The cCQS in (b) generates a qubit time series of
nonorthogonal pure states $\ket{0}\bra{0}$ and $\ket{+}\bra{+}$, where $\ket{+}
= \frac{1}{\sqrt{2}}(\ket{0}+\ket{1})$.

\paragraph*{Measured Qubit Processes}
\label{sec:MeasQProcess} 

The observer interacts with such processes by applying to each qubit a
projective measurement, consisting of the set of orthonormal measurement
operators $\{E_0, E_1\}$ with measurement basis $E_0 = \ket{\psi_0}
\bra{\psi_0}$ and $E_1 = \ket{\psi_1} \bra{\psi_1}$ parametrized by the Bloch
angles $\theta$ and $\phi$ via: \\[-15pt]
\begin{subequations}
\begin{align}
\ket{\psi_0} & = \cos{\frac{\theta}{2}}\ket{0}
  + e^{i\phi}\sin{\frac{\theta}{2}} \ket{1} ~\text{and}\\
\ket{\psi_1} & =  \sin{\frac{\theta}{2}}\ket{0}
  - e^{i\phi}\cos{\frac{\theta}{2}} \ket{1}
  ~.
\end{align}
\label{eq:param}
\end{subequations}\\[-15pt]
The outcome of each measurement can then be labeled $0$ or $1$, respectively,
resulting in a binary classical stochastic process. 

Knowledge of the controller and the measurement basis allows us to construct an
HMM describing the measured qubit process itself. This \emph{measured} cCQS has
the same states and stationary distribution $\pi$ as the original cCQS. It has
labeled transition matrices $\{T^{(\msym)}\}$ with $\msym \in \Abet$:\\[-15pt]
\begin{align}
    T^{(\msym)} = \sum\limits_{\rho_j} \CQST^{\rho_j} \Prob(\msym |\rho_j)
  ~,
\label{eq:trans_probs}
\end{align}\\[-15pt]
where $\Prob(i|\rho_j) = tr(E_i \rho_j E_i^\dagger)$ and the cCQS labeled
transition matrices $\CQST^{\rho_j}$ are defined in Fig. \ref{fig:cCQSs}. A key
step is that one can determine the HMM of the measured process by composing the
measurement operator with the the qubit controller HMM that governs the cCQS.
See the HMM in Fig. \ref{fig:cCQSs} (c). It generates the classical process
resulting from measuring the qubit process generated by Fig. \ref{fig:cCQSs}
(b) with angles $\phi = 0$ and $\theta = \pi/2$.

\paragraph*{Uncountable Predictive Features}
\label{sec:PredFeatures} 

One would hope that, since here we know the measured cCQS---e.g., shown in
Fig.\ref{fig:cCQSs}(c) for the example there---we can apply Eqs.
(\ref{eq:uni_hmu})-(\ref{eq:Cmu}) to calculate our measures directly from that
model. Unfortunately, a problem arises. The measured cCQS is not an \eM since
the generated measurement sequences are not in one-to-one correspondence with
the internal state sequences. This is the problem of cCQS \emph{nonunifilarity}
and it stymies any attempt to directly calculate cCQS randomness and structure.
In fact, and this is our first result, nonunifilarity is generic to cCQSs and
even to more general qubit sources. Thus, measurement induces nonunifilarity
and engenders, therefore, all of the resulting complications, whose
consequences we now explore in detail.

The puzzle is that we have a model in hand and it---the measured cCQS of Fig.
\ref{fig:cCQSs}(c)---generates the measured qubit process, but we cannot use it
to directly determine even the most basic process properties. What we would
like and could use is an \eM\ that represents the measured process itself.
Fortunately, the measured cCQS can be converted to an \eM\ by calculating the
cCQS's \emph{mixed states}; see SM \ref{app:MixedStates}. Here, we give a
synopsis.

As first formalized by Blackwell \cite{Blac57b}, an $N$-state HMM's mixed
states are conditional probability distributions
$\eta(\meassymbol_{-\ell:0}) = \Pr(\AlternateState_0 |
\MeasSymbol_{-\ell:0} = \meassymbol_{-\ell:0}$) over the measured HMM's
internal states $\AlternateState$ given all sequences $\meassymbol_{-\ell:0}
\in \Abet^\ell$. The collection over all of a process' allowed sequences
induces a (Blackwell) measure $\mu$ on the state distribution
$\Pr(\AlternateState)$ $N$-dimensional simplex $\MxSSet$. The mixed states
together with the mixed-state transition dynamic (see SM \ref{app:MixedStates})
give an HMM's \emph{mixed-state presentation} (MSP).

A mixed state answers the question, given that one knows the HMM structure and
has seen a particular sequence, what is the best guess of the internal state
probabilities? Transient mixed states are those state distributions after
having seen finite-$\ell$ sequences, while recurrent mixed states are those
remaining with positive probability in the limit that $\ell \to \infty$.
Recurrent mixed states exactly correspond to causal states $\CausalStateSet$
\cite{Crut08b}. When $\Cmu$ diverges, recurrent mixed states lay in an
uncountable (Cantor-like) set $S \subseteq \MxSSet$; see Fig. \ref{fig:cCQSs}(d).

To emphasize, and this is our second result, measurement-induced nonunifilarity
results in the number of causal states diverging. That is, despite the internal
controller having only a finite number of states and the controlling cCQS being
unifilar, predicting the observed process requires an uncountable number of
features in the typical case. In this way, measurement induces infinite complexity that confronts an
observer of a quantum process. It also introduces a new and fundamental
challenge: How to define and quantify the resulting randomness and complexity?

\paragraph*{Measurement-Induced Randomness and Statistical Complexity}
\label{sec:MeasurementMeasures} 

The uncountable number of mixed states also renders the complexity measure
expressions in Eqs. (\ref{eq:uni_hmu})-(\ref{eq:Cmu}) unusable. Fortunately,
Blackwell provided a formal expression for the entropy rate \cite{Blac57b} by
showing that an HMM's mixed-state presentation is unifilar. The entropy rate is
then an integral over the invariant Blackwell measure $\mu(\mxst)$ in the
mixed-state simplex $\MxSSet$:\\[-15pt]
\begin{align}
\hmu^B = - \int_{\MxSSet} d \mu(\mxst) \sum_{\msym \in \MeasAlphabet}
  \Pr(\msym |\mxst) \log_2 \Pr(\msym |\mxst)
  ~.
\label{eq:hmuB}
\end{align}\\[-15pt]
Recently, Ref. \cite{Jurg19a} introduced a constructive approach to evaluate
this integral by establishing contractivity of the simplex maps---the
substochastic transition matrices of Eq. (\ref{eq:trans_probs})---and showing
that the mixed-state process is ergodic. Rather than integrate over the
Blackwell measure $\mu(\mxst)$, such as in Fig. \ref{fig:cCQSs}(d), these
properties say that we can average over a time-series of mixed states $\mu_t$
to get the measured CQSs entropy rate:\\[-15pt]
\begin{align}
\widehat{\hmu^B} = - \lim_{\ell \to \infty} \frac{1}{\ell}
  \sum_{i = 0}^\ell \sum_{\msym \in \MeasAlphabet}
  \Pr(\msym |\mxst_i) \log_2 \Pr(\msym |\mxst_i)
  ~,
\label{eq:CQShmu}
\end{align}\\[-15pt]
where $\Pr(\msym |\mxst_i) = \mxst(\ms{0}{i}) \cdot T^{(\msym)} \cdot \One$,
$\ms{0}{i}$ is the first $i$ symbols of an arbitrarily long sequence
$\ms{0}{\infty}$ generated by the process, and $\One$ is a column-vector of all
$1$s. And so, we can now quantify the measured qubit process' randomness---the
entropy rate of the cCQS considered as a Shannon information source.

\begin{figure*}[htbp]
\hspace{0.5in}
\includegraphics[width=0.9\textwidth]{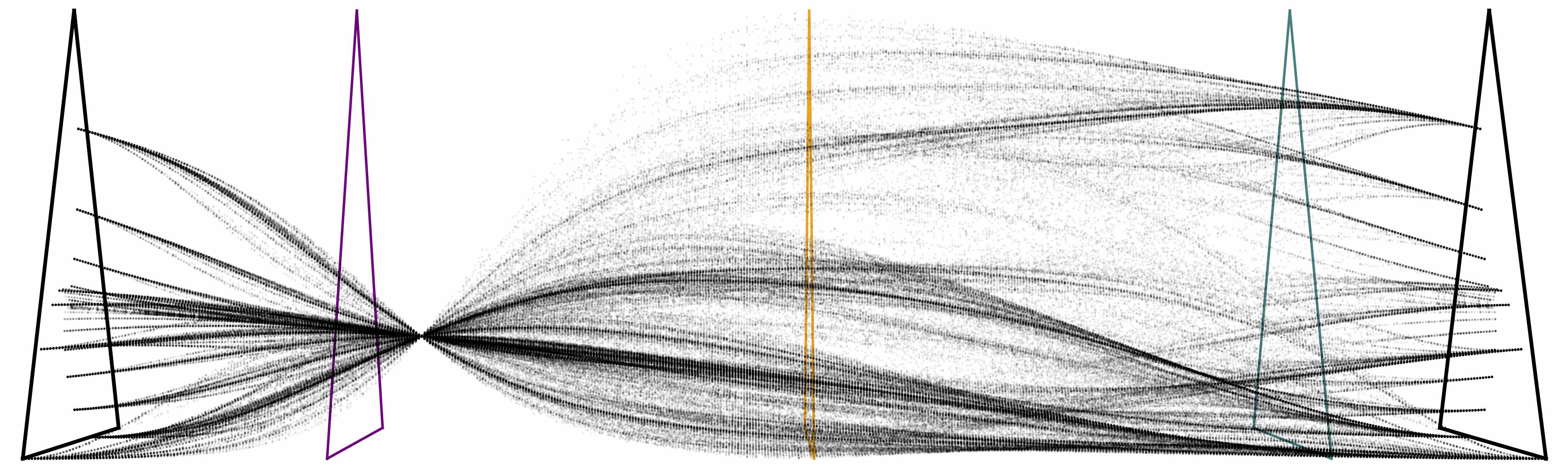}
\includegraphics[width=\textwidth]{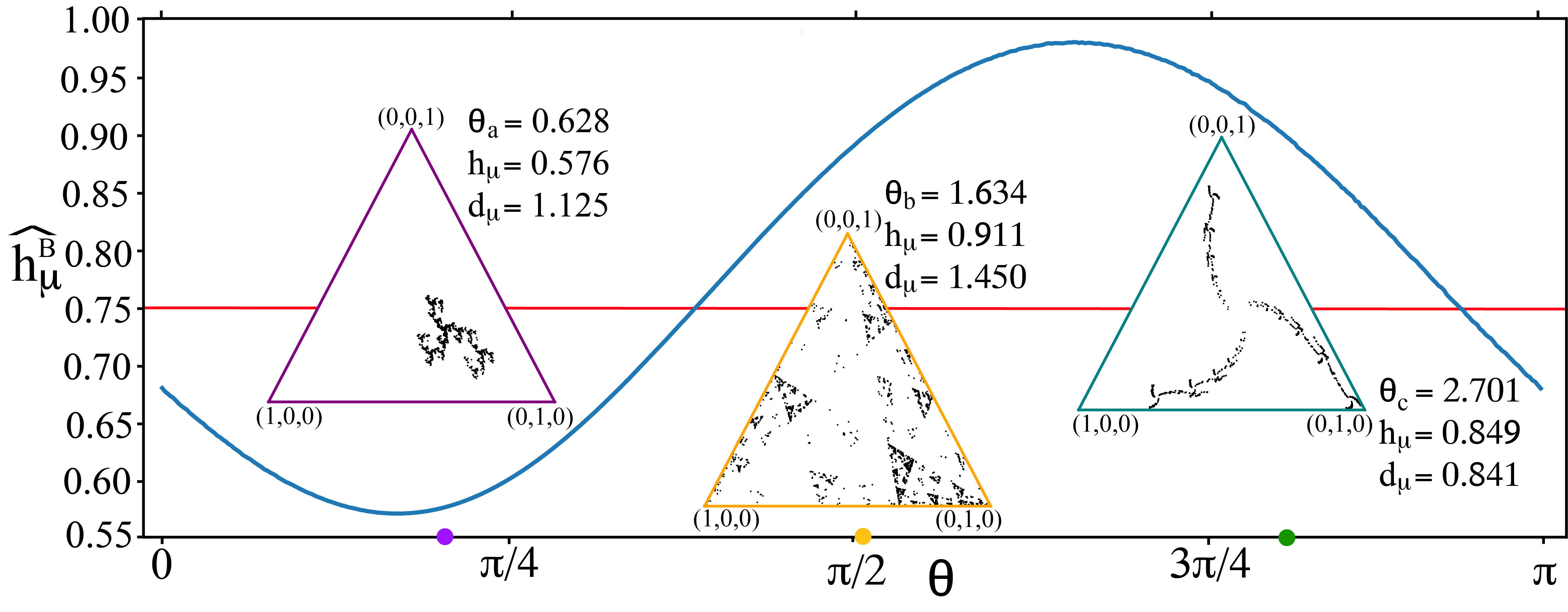}
\caption{Measurement-induced randomness and structure in a qubit process:
	(Top) Mixed state sets when measuring the qubit process
	of Fig. \ref{fig:cCQSs}(b) as a function of measurement angle $\theta \in
	[0,\pi]$.
	(Bottom) Entropy rate $\widehat{\hmu^B}$ (blue curve) as a function of
	angle. Horizontal line (red) is the entropy rate of the
	(unmeasured) qubit sequences: $\hmu^g = 3/4$ bit per output qubit.
	(Insets) Mixed states of the qubit process observed at
	three measurement angles: (a) $\theta_a = 0.628$ (purple) (green), (b)
	$\theta_b = 1.634$ (orange), and (c) $\theta_c = 2.701$. The measured
	process entropy rates $\hmu$ and statistical complexity dimensions
	$\CmuDim$ given there. Both mixed states and complexity measures 
	computed with $\ell =10^6$ iterates.
	}
\label{fig:nemote}
\end{figure*}

As a CQS generates its output process, how much memory (statistical complexity
$\Cmu$) does it use? Since the number of causal states diverges, the answer
requires care. Generically, in this case the causal state set $\CausalStateSet$
is uncountable and, visually, a rather complicated-looking self-similar set;
again see Fig. \ref{fig:cCQSs}(d). The consequence is that the amount memory, as
monitored by $\Cmu$, diverges. And so, instead, we track its rate of
divergence---the \emph{statistical complexity dimension} $\CmuDim$ of the
Blackwell measure $\mu$ on $\MxSSet$ \cite{Marz17a}:\\[-15pt]
\begin{align}
\CmuDim = \lim_{\epsilon \to 0}
  - \frac{\H_\epsilon [\AlternateState]}{\log_2 \epsilon}
  ~,
\label{eq:CmuDim}
\end{align}\\[-15pt]
where $\H_\epsilon [Q]$ is the Shannon entropy (in bits) of the
continuous-valued random variable $Q$ coarse-grained at size $\epsilon$ and
$\AlternateState$ is the random variable associated with the mixed states $\eta
\in \AlternateStateSet$. SM \ref{app:MixedStates} and \ref{app:CmuDim} develop
an upper bound on $\CmuDim$ that can be accurately determined from the measured
process' entropy rate $\widehat{\hmu^B}$ above and the mixed-state process'
Lyapunov characteristic exponent spectrum $\Lambda$. As discussed in SM
\ref{app:CmuDim}, this upper bound may be close approximation to $\CmuDim$, but
may also be a strict inequality. Which is the case can be easily determined
from the HMM's form, as well as through comparison to direct numerical
estimation of $\CmuDim$. And so, one can determine that there is a nonzero, but
finite $\Cmu$ divergence rate.

\paragraph*{Measurement Dependence}
\label{sec:MeasureDep} 

Equations (\ref{eq:param}) and (\ref{eq:trans_probs}) indicate that the choice
of measurement basis alters the observed process. This, in turn, implies that
the process entropy rate and statistical complexity dimension also depend on
measurement. To explore this with an example, we calculate the dependence of
the above complexity measures as a function of measurement angle $\theta$, with
fixed $\phi$ for the cCQS of Fig. \ref{fig:cCQSs}(b), determining the measured
cCQS at each measurement setting.

Figure \ref{fig:nemote} (top) shows the results. The cCQS is measured in $500$
different bases, holding $\phi = 0$ fixed and varying $\theta \in [0,\pi]$
uniformly. For each measured cCQS the MSP is computed and the resulting series
of mixed state sets is plotted. Figure \ref{fig:nemote} (bottom) plots
$\widehat{\hmu^B}(\theta)$ and highlights three particular measurement angles
$\{\theta_a,\theta_b,\theta_c\}$, showing the recurrent states found in
latter's mixed-state simplices. MSP entropy rate and the statistical complexity
dimension are estimated using Eqs. (\ref{eq:CQShmu}) and (\ref{eq:CmuDim}),
respectively.

Common characteristics are apparent, such as a smooth behavior of
$\hmu(\theta)$ with well-defined maxima and minima and the systematic change in
the MSP structure a function of $\theta$ which is consistent with the quoted
dimensions $\CmuDim$. Angles $\theta = 0$ and $\theta = \pi$ give particularly
simple behaviors with finite statistical complexity and $\CmuDim = 0$, in
accord with the countable MSPs there. The measured machines at these two values
of $\theta$ are identical, aside from a symbol swap---all $0$'s become $1$'s
and vice versa. As such, they both have $\Cmu = 0.6813$ bits.

Figure \ref{fig:nemote} (top) exhibits a case of interest at $\theta = \pi/4$.
The mixed states converge to a single point: a single-state machine that
represents a biased coin. This occurs since the underlying cCQS has a
binary quantum alphabet $\Abet_Q = \{\rho_0, \rho_+\}$ and the measurement
basis corresponding to $\phi = 0$ and $\theta = \frac{\pi}{4}$ with basis
vectors $\ket{\psi_0}$ and $\ket{\psi_1}$ is such that $\Pr(0|\rho_0) =
\Pr(0|\rho_+)$ and $\Pr(1|\rho_0) = Pr(1|\rho_+)$. This basis is equidistant
from both quantum states in $\Abet_Q$. Therefore, applying the measurement to
one state or the other yields the same probability distribution over outcomes.
One loses all information about the underlying structure and the measured cCQS
generates an \emph{independent identically distributed} process.

To compare the randomness and organization of the underlying generator process,
the horizontal line in the $\widehat{\hmu^B}(\theta)$ plot gives the entropy
rate of the (unmeasured) qubit process: $\hmu^g = 3/4$ bit per output qubit.
Its statistical complexity is $\Cmu^g = \H[\pi] = 1.5$ bits. The differences
between these constant values and those of the measured cCQS values makes it
clear that quantum measurement can both add or remove randomness and structure.

\vspace{-0.05in}
\paragraph*{Conclusion}
\label{sec:Conclusion} 
That randomness and complexity arise when observing qubit processes can be too
facilely appreciated. Indeed, quantum measurement often comes steeped in
mystery. We dispelled some of that mystery by showing that (i) an infinite
number of predictive features are required to describe measured qubit time
series and (ii) measurement both introduces and subtracts information and
correlation. These characters of measurement greatly complicate learning about
the informational and dynamical organization of quantum systems. However, at
least now, we can appreciate more fully what the task is, what mechanism drives
it (nonunifilarity), and why it is challenging.

Precisely stating how this occurs and quantitatively identifying what is added
and removed was nontrivial, however, requiring several innovations. We needed
to introduce new methods to estimate qubit process randomness
(Shannon-Kolmogorov-Sinai entropy rate). We then had to introduce a wholly new
quantity---the statistical complexity dimension---to track stored information
and memory resources. And, to constructively work with these quantities
required extending results from ergodic theory, abstract dynamical systems, and
information theory to cCQSs. Thus, analyzing the quantum physics led us to
introduce novel theory and efficient algorithms for quantifying the randomness
and complexity of ergodic, stationary processes generated by nonunifilar hidden
Markov models. Mathematically, these gave a constructive answer to the
longstanding information-theoretic problem of identifying functions of Markov
chains---a problem that until now had only been formally, not constructively,
solved \cite{Blac57b}.

\vspace{-0.05in}
\paragraph*{Acknowledgments}
\label{sec:acknowledgments}
We thank Cina Aghamohammadi, Fabio Anza, Sam Loomis, Sarah Marzen, and Todd
Pittman for helpful discussions. AMJ, AEVL, and JPC thank the Santa Fe Institute
and JPC thanks the Telluride Science Research Center, Institute for Advanced
Study at the University of Amsterdam, and California Institute of Technology
for their hospitality during visits. This material is based upon work supported
by, or in part by, U. S. Army Research Laboratory and the U. S. Army
Research Office under contracts W911NF-13-1-0390 and W911NF-18-1-0028.

\vspace{-0.05in}
\paragraph*{Supplementary Materials}
Materials and Methods: Reviews classical and quantum stochastic processes,
information measures, and mixed states and provides details of the numerical
calculations.

\onecolumngrid
\clearpage
\begin{center}
\large{Supplementary Materials}\\
\vspace{0.1in}
\emph{\ourTitle}\\
\vspace{0.1in}
{\small
Ariadna Venegas-Li, Alexandra Jurgens, and James P. Crutchfield
}
\end{center}

\setcounter{equation}{0}
\setcounter{figure}{0}
\setcounter{table}{0}
\setcounter{page}{1}
\makeatletter
\renewcommand{\theequation}{S\arabic{equation}}
\renewcommand{\thefigure}{S\arabic{figure}}
\renewcommand{\thetable}{S\arabic{table}}

The Supplementary Materials review the quantum and classical formalisms with
which we work and present details, derivations, and explanations for the theoretical claims and simulation results.

\section{Quantum Formalism}
\label{qpp:QuantumFormalism}

The quantum evolution of a qubit is typically considered to occur in the
Hilbert space $H_2$ which contains the one-parameter (time) family of states
$\rho(t)$. While this representation is appropriate for many problems, it does
not accurately describe the time series of qubits that concern us. In our time
series, a different qubit is emitted at each time step $t$. The time parameter
$t$ is discrete and labels the qubit state $\rho_t$ emitted at time $t$. A
different Hilbert space $H_2^t$ contains the state of each distinct qubit---the
state of the qubit emitted at time $t$ belongs to the Hilbert space: $\rho_t
\in H_2^t$. The state of the entire bi-infinite time series
lies in the Hilbert space is $\mathcal{H} = \lim_{\ell \rightarrow \infty}
\bigotimes_{t = -\ell}^{+\ell} H_2^t$. And, when considering a finite part of
the time series of length $\ell$, the Hilbert space of interest is a truncation
of $\mathcal{H}$ denoted $\mathcal{H}_{t:t+l} = \bigotimes_{k = t}^{t+l-1}
H_2^k$.

In our setting, a \emph{classically controlled qubit source} (cCQS) emits a
qubit at each time step. The cCQS also determines state $\rho_t$ of each output
qubit. The latter is taken to be a pure state and it remains constant until
measured. The qubit chain's state then is $\ldots \rho_{t-1} \otimes \rho_t
\otimes \rho_{t+1} \otimes \ldots$. These restrictions guarantee that there is
no temporal entanglement and that one can apply a single-qubit projective
measurement $E$ to each output qubit without affecting the states of the other
qubits in the time series.

\begin{figure}[htbp]
\centering
\includegraphics[width=0.4\textwidth]{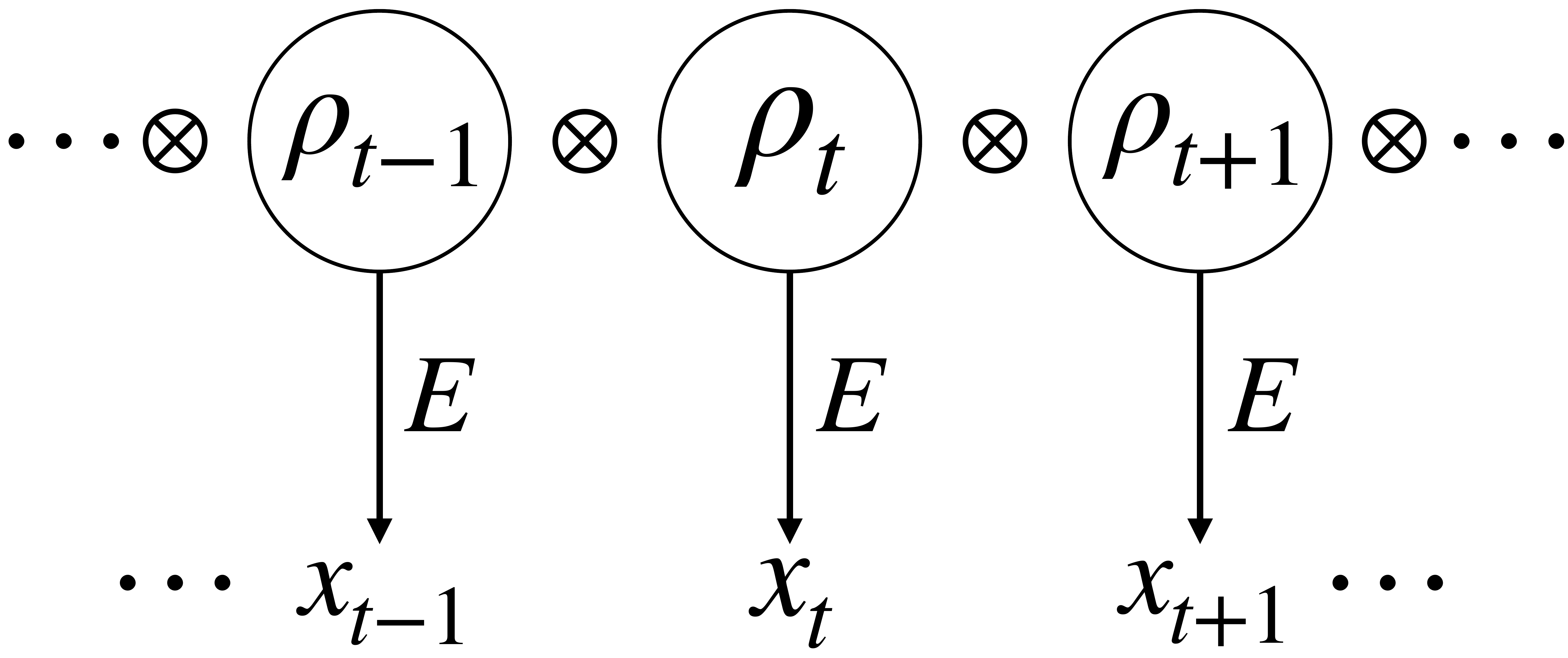}
\caption{A controlled qubit source emits a discrete-time stochastic process
	of qubits in pure states $\rho_t$. Applying measurement $E$ to the qubit
	emitted at time $t$ in state $\rho_t$ yields outcome $x_t$. Measuring
	each qubit at each time step yields a classical stochastic process.
	}
\label{fig:proc_diag}
\end{figure}

We define measurement operators $E_0 = \ket{\psi_0}\bra{\psi_0}$ and $E_1 =
\ket{\psi_1}\bra{\psi_1}$ with their basis vectors parametrized as in Eq.
(\ref{eq:param}). As depicted in Fig. \ref{fig:proc_diag}, applying measurement
$E$ to each qubit $\rho_t$ yields a classical time series realization $\ldots
x_{t-1} x_t x_{t+1}\ldots$ which depends on both the internal quantum
process and the projective measurement $E$. We concentrate on quantifying
randomness and structure in these observed classical stochastic processes, as
they capture the classically-observed statistical properties of the internal
quantum process.

The developments here complement recent progress on representing classical
processes via quantum channels, which showed that quantum representations can
be markedly smaller than classical \cite{Gu12a, Maho15a, Riec15b} and that
classical and quantum physics are at odds when it comes to measures of
organization \cite{Agha16a,Loom18a}. In a similar, complementary way, we hope
that our results aid in developing a systematic description of memoryful quantum
processes built on experimentally accessible quantities \cite{Poll18a}.

\section{Classical Formalism and Methods}
\label{app:ClassicalFormalism}

First, we recall notation for classical stochastic processes and their HMM
generators, explicitly calling out the key subclass of unifilar HMMs.  The
sections then turn to information theory and computational mechanics for
stationary, ergodic processes. They review mixed-state presentations and
measures of randomness and structure which are needed to characterize processes
emitted by nonunifilar generators. And, they end outlining further explorations
that take into account the several disparate fields involved.

\subsection{Stochastic Processes}
\label{app:processes}

As many of the tools used here come from the theory of classical stochastic
processes, we introduce several definitions and notation for the reader less
familiar with it. A classical stochastic process $\bf{\MSym}$ is a series of
random variables and a specification of the probabilities of their
realizations. The random variables corresponding to the behaviors are denoted
by capital letters $\ldots \MSym_{t-2}, \MSym_{t-1}, \MSym_t, \MSym_{t+1},
\MSym_{t+2} \ldots$. Their realizations are denoted by lowercase letters
$\ldots \msym_{t-2}, \msym_{t-1}, \msym_t, \msym_{t+1}, \msym_{t+2} \ldots$,
with the $\msym_t$ values are drawn from a discrete alphabet $\Abet$. Blocks
are denoted as: $\MSym_{t: t+l} = \MSym_t, \MSym_{t+1}, \ldots \MSym_{t+l-1}$,
the left index is inclusive and the right one exclusive.

For our purposes, we consider \emph{stationary} stochastic processes, in which
the probability of observing behaviors is time-translation invariant:
\begin{align*}
\Prob(X_{t:t+\ell} = x_{t:t+\ell}) = \Prob(X_{0:\ell} = x_{0:\ell})
  ~,
\end{align*}
for all $t$ and $\ell$.
We concentrate, in particular, on processes that can be generated by a hidden
Markov model.

\subsubsection{Hidden Markov Models and Unifilarity}
\label{subsec:hmmuni}

A \emph{hidden Markov model} (HMM) is a quadruple $(\SSet,
\Abet, \{T^\msym\},\pi)$ consisting of: 
\begin{itemize}
\item $\SSet$ is the set of hidden states. 
\item $\Abet$ the alphabet of symbols that the HMM emits on state-to-state
	transitions at each time step. 
\item $\{T^\msym: \msym \in \Abet\}$  is the set of labeled transition matrices
	such that $T^\msym_{ij} = \Prob(\msym, \cs_j|\cs_i)$ with $\cs_i, \cs_j \in
	\SSet$. That is, $T^\msym_{ij}$ denotes the probability of the HMM
	transitioning from state $\cs_i$ to state $\cs_j$ while emitting symbol
	$\msym$. 
\item $\pi$ is the stationary state distribution determined from the left
	eigenvector of $T = \sum_{\msym \in \Abet} T^\msym$ normalized in
	probability. 
\end{itemize}

Figures \ref{fig:cCQSs} (a), (b), and (c) are three-state HMMs over qubit alphabets
$\mathcal{A}_Q = \{\rho_0 = \ket{0}\bra{0}, \rho_1 = \ket{1}\bra{1}\}$ and
$\mathcal{A}_Q = \{\rho_0 = \ket{0}\bra{0}, \rho_1 = \ket{+}\bra{+}\}$, and
over binary alphabet $\mathcal{A} = \{0,1\}$, respectively.

An HMM property that proves to be essential is unifilarity. An HMM is
\emph{unifilar} if, given a hidden state, the emitted symbol $\msym \in \Abet$
uniquely identifies the next state. Equivalently, for each labeled transition
matrix $T^\msym$, there is at most one nonzero entry in each row. Unifilarity
ensures that a sequence of emitted symbols has a one-to-finite correspondence
with sequences of hidden-state paths. The HMMs in Figs. \ref{fig:cCQSs} (a) and
(b) are unifilar.

In contrast, if an HMM is nonunifilar the set of allowed hidden-state state
paths corresponding to a sequence of emitted symbols grows exponentially with
sequence length. The HMM in Fig. \ref{fig:cCQSs} (c) is nonunifilar. This can
be easily checked by noting that state sequences $AAA$, $AAB$, and $ABC$ all
emit the single output $000$. Most basically, nonunifilarity makes inferring
the underlying states and transitions from the generated output process a
difficult task.

\subsection{Measures of Complexity}
\label{app:moc}

We consider two complexity measures that have clear operational meanings:
a process' intrinsic randomness and the minimal memory resources required to
predict its behavior.

The intrinsic randomness of a classical stochastic process $\bf{\MSym}$ is
measured by its \emph{entropy rate} \cite{Cove91a}:
\begin{align}
\hmu = \lim\limits_{\ell \rightarrow \infty} \frac{H(\ell)}{\ell}
  ~,
\label{eq:hmu1}
\end{align}
where $H(\ell) = \H[\Prob(\MS{0}{\ell-1})]$ is the Shannon entropy for
length-$\ell$ blocks. That is, a process' intrinsic randomness is the
asymptotic average Shannon entropy per emitted symbol---a process' entropy
growth rate.

Shannon showed that this is the same as the asymptotic value of the entropy of
the next symbol conditioned on the past \cite{Shan48a}:
\begin{align}
\hmu = \lim\limits_{\ell \rightarrow \infty} \H[\MSym_0 | \MS{-\ell}{0}]
  ~.
\label{eq:hmu2}
\end{align}
This can be interpreted as how much information is gained per measurement once all the possible structure in the sequence has been captured. 

Determining $\hmu$ is possible only for a small subset of
stochastic processes. Shannon \cite{Shan48a} gave closed-form expressions for
processes generated by Markov Chains (MC), which are ``unhidden'' HMMs---they
emit their states as symbols. Making use of Eq.
(\ref{eq:hmu2}), he proved that for MC-generated processes the entropy rate is
simply the average uncertainty in the next state: 
\begin{align}
\hmu =  - \sum\limits_{\cs \in \CausalStateSet} \Prob(\cs)
\sum\limits_{\cs^\prime \in \CausalStateSet} T_{\cs \cs^\prime} \log
T_{\cs \cs^\prime}
  ~,
\label{eq:mc_hmu}
\end{align}
where $T$ is the MC's transition matrix and $\CausalStateSet$ its set of states.

Another special case for which the entropy rate can be exactly computed is for
processes generated by unifilar HMMs (uHMMs) \cite{Cove91a}. This class
generates an exponentially larger set of processes than possible from MCs.
Since each infinite sequence of emitted symbols corresponds to a unique
sequence of internal states, or at most a finite number, the process entropy
rate is that of the internal MC. And so, one (slightly) adapts Eq.
(\ref{eq:mc_hmu}) to calculate $\hmu$ for these processes. The expression is
presented in Eq. (\ref{eq:uni_hmu}).

A process' structure is most directly analyzed by determining its minimal
predictive presentation, its $\eM$. A simple measure of structure is then given
by the number of causal states $|\CausalStateSet|$ or by the \emph{statistical
complexity} $\Cmu$ defined in Eq. (\ref{eq:Cmu}), which is the Shannon
information $\H[\CausalStateSet]$ stored in the causal states. Since the set of
causal states is minimal, $\Cmu$ measures of how much memory about the past a
process remembers. Said differently, $\Cmu$ quantifies the minimum amount of
memory necessary to optimally predict the process' future.

However, for processes generated by nonunifilar HMMs---such as generic measured
cCQSs---both $\hmu$ and $\Cmu$ given by Eqs. (\ref{eq:uni_hmu}) and
(\ref{eq:Cmu}) are incorrect. The former overestimates the generated process'
$\hmu$, since uncertainty in the next symbol is not in direct correspondence
with the uncertainty in the next internal state. In fact, there is no exact
general method to compute the entropy rate of a process generated by a generic
nonunifilar HMM. One has only the formal expression of Eq. (\ref{eq:hmuB})
which refers to a abstract measure that, until now, was not constructively
determined. For related reasons, the statistical complexity $\Cmu$ given by Eq.
(\ref{eq:Cmu}) applied to that abstract measure is useless---it simply
diverges.

For processes generated by nonunifilar HMMs one can take a very pragmatic
approach to estimate randomness and structure from process realizations
(measured or simulated time series) using information measures for sequences of
finite-length ($\ell$), such as reviewed in Refs. \cite{Crut01a,Jame11a}. This
approximates the sequence statistics as an order-$\ell$ Markov process. The
associated conditional distributions capture only finite-range correlations,
becoming: $\Pr(\MeasSymbol_{t:\infty} | \meassymbol_{-\infty:t}) =
\prod_{i=t}^{\infty} \Pr(\MeasSymbol_i | \MeasSymbol_{i-\ell} \ldots
\MeasSymbol_{i-1})$. This approach is data-intensive and the complexity
estimators have poor convergence.

Addressing the shortcomings for processes generated by nonunifilar HMMs
requires introducing the fundamental concepts of predictive features and a
process' mixed-state presentation.

\subsection{Calculating Mixed States}
\label{app:MixedStates}

The finite Markov-order approach seems to make sense empirically. However, one
would hope that, if we know the nonunfilar HMM and therefore have a model
(states and transitions) that generates the process at hand, we can calculate
randomness and structure directly from that model or, at least, do better than
by using slowly-converging order-$\ell$ Markov approximations. The approach is
to construct a unifilar HMM---the process' \eM---from the the nonunfilar HMM.
This is done by calculating the latter's \emph{mixed states}.

Each mixed state tracks the probability distribution over the the nonunifilar HMM's internal states, conditioned on the possible sequences of observed symbols. In other words, the mixed states represent \emph{states of knowledge} of the nonunifilar HMM's internal states. This also allows one to compute the transition dynamic between mixed states, forming a unifilar model for the same process as generated by the original nonunifilar HMM.

Explicitly, assume that an observer has an HMM presentation $M$ for a process
$\Process$, and before making any observations has a probabilistic knowledge of
the current state---the state distribution  $\mxst_0 = \Prob(\CausalState)$.
Typically, prior to observing any system output the best guess is $\mxst_0 =
\pi$.

Once $M$ generates a length-$\ell$ word $w = \msym_{0} \msym_{1} \ldots
\msym_{\ell-1}$ the observer's \emph{state of knowledge} of $M$'s current state
can be updated to $\mxst(w)$, that is: 
\begin{align}
\mxst_\cs(w) & \equiv \Prob(\CausalState_\ell = \cs |
  \MS{0}{\ell}=w, \CausalState_0 \sim \pi)
  ~.
\label{eq:MixedState}
\end{align}
The collection of possible \emph{states of knowledge} $\mxst(w)$ form the set
$\MxSSet$ of $M$'s \emph{mixed states}:
\begin{align*}
\MxSSet = \{ \mxst(w): w \in \MeasAlphabet^+, \Pr(w) > 0 \}
  ~.
\end{align*}
And, we have the mixed-state measure $\MxSMeasure(\mxst)$---the probability of
being in a mixed state:
\begin{align*}
\Pr(\mxst(w)) & = \Prob(\CausalState_\ell |
  \MS{0}{\ell}=w, \CausalState_0 \sim \pi) \Pr(w)
  ~.
\end{align*}

From this follows the probability of transitioning from $\mxst(w)$ to
$\mxst(w\msym)$ on observing symbol $\msym$:
\begin{align*}
\Pr(\mxst(w\msym) | \mxst(w))
  & = \Pr(\msym|\CausalState_\ell \sim \mxst(w))
  ~.
\end{align*}
This defines the mixed-state dynamic in terms of the original process, not in
terms of an HMM presentation of the latter. Together the mixed states and their
dynamic give the HMM's \emph{mixed state presentation} (MSP) $\MSP = \{\MxSSet,
\MxSDyn \}$ \cite{Blac57b}. 

Given an HMM presentation, though, 
we can explicitly calculate its MSP. The probability of generating
symbol $\msym$ when in mixed state $\mxst$ is:
\begin{align}
\Pr(\msym |\mxst) = \mxst \cdot T^{(\msym)} \cdot \One
  ~,
\label{eq:SymbolFromMixedState}
\end{align}
with $\One$ a column vector of $1$s. Upon seeing symbol $\msym$, the
current mixed state $\mxst_t$ is updated:
\begin{align}
\mxst_{t+1}(x) = \frac{\mxst_t \cdot T^{(\msym)}}
  { \mxst \cdot T^{(\msym)} \cdot \One }
  ~,
\label{eq:MxStUpdate}
\end{align}
with $\mxst_0 = \mxst(\lambda) = \pi$ and $\lambda$ the null sequence.

Thus, given an HMM presentation we can calculate the mixed state of Eq.
(\ref{eq:MixedState}) via:
\begin{align*}
\mxst(w) = \frac{\pi \cdot T^{(w)}}{\pi \cdot T^{(w)} \cdot \One}
  ~.
\end{align*}
The mixed-state transition dynamic is then:
\begin{align*}
\Pr(\mxst_{t+1},\msym|\mxst_t) & = \Pr(\msym|\mxst_t) \\
   & = \mxst_t \cdot T^{(\msym)} \cdot \One
   ~,
\end{align*}
since Eq. (\ref{eq:MxStUpdate}) tells us that, by construction, the MSP is
unifilar. That is, the next mixed state is a function of the previous and the
emitted (observed) symbol.

Now, with a unifilar presentation one is tempted to directly apply Eqs.
(\ref{eq:uni_hmu}) and (\ref{eq:Cmu}) to compute measures of randomness and
structure, but another challenge prevents this. With a small number of
exceptions, the MSP of a process generated by a nonunifilar HMM has an
uncountable infinity of states $\mxst$ \cite{Marz17a}. Practically, this means
that one cannot construct the full MSP, that direct application of
Eq. (\ref{eq:uni_hmu}) to compute the entropy rate is not feasible, and that
$|\CausalStateSet|$ diverges and, typically, so does $\Cmu$.

\subsection{Entropy Rate of Nonunifilar Processes}
\label{app:HmuBlackwell}

Fortunately, when working with ergodic processes, such as those addressed here, one can accurately estimate the MSP by generating a word $w_\ell$ of sufficiently long length \cite{Jurg19a}. The main text addresses in some detail how to use this to circumvent the complications of uncountable mixed states when computing the entropy rate. Specifically, with the mixed states in hand computationally, accurate numerical estimation of the entropy rate of a process generated by a nonunifilar HMM is given by using the temporal average specified in Eq. (\ref{eq:CQShmu}). The development of that expression and the proof that it is correct is given in Ref. \cite{Jurg19a}.

This handily addresses accurately estimating the entropy rate of nonunifilar
processes. And so, we are left to tackle the issue of these process' structure
with the \emph{statistical complexity dimension.} This requires a deeper
discussion.

\subsection{Statistical Complexity Dimension}
\label{app:CmuDim}

$\Cmu$ diverges for processes generated by generic HMMs, as they are
nonunifilar with probability one and that, in turn, leads to an uncountable
infinity of mixed states. To quantify these processes' memory resources
one tracks the rate of divergence---the \emph{statistical complexity dimension}
$\CmuDim$ of the Blackwell measure $\mu$ on $\MxSSet$:
\begin{align}
\CmuDim = \lim_{\epsilon \to 0}
  - \frac{\H_\epsilon [\AlternateState]}{\log_2 \epsilon}
  ~,
\end{align}
where $\H_\epsilon [Q]$ is the Shannon entropy (in bits) of the
continuous-valued random variable $Q$ coarse-grained at size $\epsilon$ and
$\AlternateState$ is the random variable associated with the mixed states
$\eta \in \AlternateStateSet$.

$\CmuDim$ is determined by the measured process' entropy rate
$\widehat{\hmu^B}$, as given by Eq. (\ref{eq:CQShmu}), and the mixed-state
process' \emph{spectrum of Lyapunov characteristic exponents} (LCEs). The
latter is calculated from Eq. (\ref{eq:trans_probs})'s labeled transition
matrices which map the mixed states $\eta_t \in \MxSSet$ according to Eq.
(\ref{eq:MxStUpdate}). The LCE spectrum $\Lambda = \{ \lambda_1, \lambda_2,
\ldots, \lambda_N: \lambda_i \geq \lambda_{i+1} \}$ is determined by
time-averaging the contraction rates along the $N$ eigendirections of this
map's Jacobian. The statistical complexity dimension is then bounded by a
modified form of the LCE dimension \cite{Fred83a}:
\begin{align}
\CmuDim \leq \LCEDim
  ~,
\end{align}
where:
\begin{align}
\LCEDim = k - 1 + \frac{\widehat{\hmu^B}
  + \sum_{i=1}^k \lambda _i}{|\lambda_{k+1}|}
\label{eq:LCEDim}
\end{align}
and $k$ is the greatest index for which $\widehat{\hmu^B} + \sum_{i=1}^k
\lambda_k > 0$. Reference \cite{Jurg19a} introduces this bound for an HMM's
statistical complexity dimension, interprets the conditions required for its
proper use, and explains in fuller detail how to calculate an HMM's LCE
spectrum.

In short, the set of mixed states generated by a generic HMM is equivalent to
the fractal set defining the attractor of a nonlinear, \emph{place-dependent
iterated function system (IFS)}. Exactly calculating dimensions---say,
$\CmuDim$---of such sets is known to be difficult. This is why here we adapt
$\LCEDim$ to iterated function systems. The estimation is conjectured to be
accurate in ``typical systems'' \cite{Kapl79a,Feng09}. Even so, in certain
cases where the IFS does not meet the \emph{open set condition}
\cite{Feng09}---for example, the inset in Fig. \ref{fig:nemote} at $\theta_b =
1.634$---the relationship becomes an inequality: $\CmuDim < \LCEDim$. This
case, which is easily detected from an HMM's form, is discussed in more detail
in Ref. \cite{Jurg19a}. 

That these subtleties require care is borne out by numerical checks on the
estimated dimensions. We estimated the box-counting dimension $\BoxDim$
\cite{Falc90a}, a finite-$\epsilon$ approximation to $\CmuDim$, for the inset
MPSs in Fig. \ref{fig:nemote}: for $\theta_a = 0.628$ $\BoxDim = 1.12$, for $\theta_b = 1.634$ $\BoxDim = 1.18$, and for $\theta_c =
2.701$ $\BoxDim = 0.85$. Thus, for measurement angles $\theta_a$ and
$\theta_c$, there is good agreement between $\BoxDim$ and the $\CmuDim$ bound.
For angle $\theta_b$, the estimated $\CmuDim$ bound is greater than the
box-counting dimension; but this is as expected. To be clear, the qualitative
conclusions that we draw about randomness and complexity in quantum measurement
do not depend on these differences in estimation. However, we do see a time
when experimentally exploring these results demands refined accuracy.

Let's close by noting that the LCE spectrum's advantage in estimating $\CmuDim$
is threefold. First, LCEs are computationally faster and cheaper to numerically
calculate than the box-counting dimension. Second, we conjecture that $\CmuDim$
of the attractors in Fig. \ref{fig:nemote} varies smoothly with angle in general.
We verified that their $\LCEDim$ does. While the box-counting algorithm also
finds this behavior, $\BoxDim$'s dependence on box shape and position in
these algorithms induces substantial estimation errors when sweeping through a
Cantor-set family. Third, the difference between $\LCEDim$ and the true
$\CmuDim$ is meaningful, as it captures the ``compressibility'' of a process
with uncountably infinite states. This makes the $\LCEDim$ bound on $\CmuDim$
useful beyond merely its role as a dimension estimator.

\end{document}